\documentclass{PoS}

\newcommand{\U}{\mathrm{U}\,}
\newcommand{\SU}{\mathrm{SU}\,}
\newcommand{\Z}{\mathbb{Z}}
\newcommand{\fen}{\mathcal{F}_{\rm eff}}

\title{
QCD Phase Diagram with 2-flavor Lattice Fermion
Formulations\thanks{RIKEN-MP-57}}
\ShortTitle{QCD phase diagram with 2-flavor lattice fermion formulations}

\author{\speaker{Taro Kimura}\\%\thanks{A footnote may follow.}\\
	Mathematical Physics Laboratory, RIKEN Nishina Center\\
        E-mail: \email{tkimura@ribf.riken.jp}}

\author{Tatsuhiro Misumi\\
        Brookhaven National Laboratory\\
        E-mail: \email{tmisumi@bnl.gov}}

\author{Akira Ohnishi\\
        Yukawa Institute for Theoretical Physics, Kyoto University\\
        E-mail: \email{ohnishi@yukawa.kyoto-u.ac.jp}}

\abstract{
We propose a new framework for investigating two-flavor lattice QCD with
finite temperature and density by applying the Karsten-Wilczek lattice fermion,
in which a species-dependent imaginary chemical potential can reduce the
number of species to two without losing chiral symmetry. 
This lattice discretization is useful for study on finite-$(T,\mu)$ QCD
since its discrete symmetries are appropriate for the case. 
To show its applicability, we study strong-coupling lattice QCD with
temperature and chemical potential. 
We derive the effective potential of the scalar meson field and obtain a
critical line of the chiral phase transition, which is qualitatively
consistent with the phenomenologically expected phase diagram. 
}

\FullConference{ The XXX International Symposium on Lattice Field Theory - Lattice 2012\\
June 24-29, 2012\\
Cairns, Australia}

\begin{document}

\section{Introduction}\label{sec:introduction}

Lattice QCD has played an important role in study of the
non-perturbative aspects of QCD.
However, its application to the finite density system has not been
established due to serious difficulty of the sign problem.
In this report we propose a new framework of investigating the 2-flavor
QCD with finite temperature and density by using the Karsten-Wilczek
(KW) lattice fermion~\cite{Karsten:1981gd%,Wilczek:1987kw
}, which
possesses only two species doublers, i.e. minimally doubled fermion.
This lattice formulation lifts degeneracy of 16 species
without breaking its chiral symmetry by introducing a species-dependent
imaginary chemical potential, instead of a species-dependent mass term
introduced in the Wilson fermion formalism.
Because of the chemical potential term, its
discrete symmetry is not sufficient to be applied to
fully Lorentz symmetric system, i.e. zero temperature and density, but
enough to study the in-medium QCD.
To show the usefulness of the KW fermion, we study strong-coupling
lattice QCD with temperature and density.

\section{Symmetry of KW-type minimally doubled fermion}

The KW fermion is a kind of minimally doubled fermions, involving only
two species doublers by introducing a species-dependent imaginary
chemical potential, which we call ``flavored chemical potential".
This term preserves its chiral symmetry and
ultra-locality~\cite{Misumi:2012uu}, but breaks some of discrete spacetime
symmetries~\cite{Bedaque:2008xs,%Bedaque:2008jm,
Kimura:2009qe%,Kimura:2009di
}.
We then need three counter terms to take a correct Lorentz
symmetric continuum limit: dimension-3, $\bar\psi i \gamma_4
\psi=i\psi^\dag \psi$,
and dimension-4 terms, $\bar\psi\gamma_4\partial_4\psi$,
$F_{j4}F_{j4}$~\cite{Capitani:2009yn%,Capitani:2010nn
}.
The fermionic part of KW fermion action with the counter terms is given by
\begin{eqnarray}
  S_{\rm KW} & = & \sum_x
  \Bigg[
   \frac{1}{2} \sum_{\mu=1}^4 \bar\psi_x \gamma_\mu
   \left(
    U_{x,x+\hat{\mu}} \psi_{x+\hat{\mu}}
    - U_{x,x-\hat{\mu}} \psi_{x-\hat{\mu}}
   \right)
   + i \frac{r}{2} \sum_{j=1}^3 \bar\psi_x \gamma_4
   \left(
    2 \psi_x 
    - U_{x,x+\hat{j}} \psi_{x+\hat{j}}
    - U_{x,x-\hat{j}} \psi_{x-\hat{j}}
   \right)
   \nonumber \\
 & &  \qquad
  + i \mu_3 \bar \psi_x \gamma_4 \psi_x
  + \frac{d_4}{2} \bar\psi_x \gamma_4
   \left(
    U_{x,x+\hat{4}} \psi_{x+\hat{4}}
    - U_{x,x-\hat{4}} \psi_{x-\hat{4}}
   \right)
  \Bigg] .
  \label{KW_action}
\end{eqnarray}
The second term in the first line including $i\gamma_4$ 
is the flavored chemical potential term, which we also call Karsten-Wilczek(KW) term.
We here introduce a parameter $r$ in analogy to Wilson fermion.
$\mu_3$ and $d_4$ are parameters for the dimension-3 and dimension-4
counter terms, respectively.
The corresponding Dirac operator in the momentum space yields
\begin{equation}
 a D_{\rm KW}(p) = 
  i \sum_{\mu=1}^4 \gamma_\mu \sin a p_\mu
  + i r \gamma_4 \sum_{j=1}^3 (1 - \cos a p_\mu)
  + i \mu_3 \gamma_4 + i d_4 \gamma_4 \sin a p_4 ,
\end{equation}
which has only two zeros at
$\bar{p}=(0,0,0,\frac{1}{a}\mathrm{arcsin}\left(-\frac{\mu_3}{1+d_4}\right))$
when $-1-d_4<\mu_3<1+d_4$ with $r=1$.
When we expand the Dirac operator around the zeros, its dispersion
relation is not Lorentz symmetric.
As shown in Ref.\cite{Misumi:2012uu}, the tuning condition for the correct dispersion relation
is given by $(1+d_4)^2=1+\mu_3^2$ at the tree level.
Moreover, it is shown that $\mu_3$ has to be tuned to control imaginary chemical
potential in $\mathcal{O}(1/a)$.

Symmetries of the lattice action (\ref{KW_action}) are chiral symmetry, 
cubic symmetry corresponding to
permutation of spatial three axes, CT and
P~\cite{Bedaque:2008xs%,Bedaque:2008jm
}:
\begin{enumerate}
 \item $\U(1)$ chiral symmetry
       ($\gamma_5\otimes\tau_3$~\cite{Creutz:2010bm,Tiburzi:2010bm,Drissi:2011da%,Drissi:2011bh
})
 \item Cubic symmetry
 \item CT
 \item P
\end{enumerate}
It is notable that these symmetries are the same as those of the finite-density lattice QCD:
As an example, we look into the naive lattice action with chemical potential, which is given by
\begin{equation}
 S_{\rm naive} = \frac{1}{2} \sum_x
  \left[
   \sum_{j=1}^3 
   \bar\psi_x \gamma_j
   \left( U_{x,x+\hat{j}} \psi_{x+\hat{j}}
        - U_{x,x-\hat{j}} \psi_{x-\hat{j}}
   \right)
   + \bar\psi_x \gamma_4
   \left( e^{\mu} U_{x,x+\hat{4}} \psi_{x+\hat{4}}
        - e^{-\mu} U_{x,x-\hat{4}} \psi_{x-\hat{4}}
   \right)
  \right] .
  \label{naive_action}
\end{equation}
The 4th direction hopping term, involving chemical potential, breaks the
hypercubic symmetry into the spatial cubic symmetry, and also C, P, and
T into CT and P, which are the same symmetries of (\ref{KW_action}).
It means that, even if we introduce chemical potential as
(\ref{naive_action}) to KW fermion, the symmetries are unchanged.  
The KW fermion with the exponential form chemical potential is given by,
\begin{eqnarray}
  S_{\rm KW} & = & \sum_x
  \Bigg[
   \frac{1}{2} \sum_{j=1}^3 \bar\psi_x \gamma_j
   \left(
    U_{x,x+\hat{j}} \psi_{x+\hat{j}}
    - U_{x,x-\hat{j}} \psi_{x-\hat{j}}
   \right)
   + i \frac{r}{2} \sum_{j=1}^3 \bar\psi_x \gamma_4
   \left(
    2 \psi_x 
    - U_{x,x+\hat{j}} \psi_{x+\hat{j}}
    - U_{x,x-\hat{j}} \psi_{x-\hat{j}}
   \right)
   \nonumber \\
 & &  \qquad
  + \frac{1+d_4}{2} \bar\psi_x \gamma_4
   \left(
    e^\mu U_{x,x+\hat{4}} \psi_{x+\hat{4}}
    - e^{-\mu} U_{x,x-\hat{4}} \psi_{x-\hat{4}}
   \right)
  + i \mu_3 \bar \psi_x \gamma_4 \psi_x
  \Bigg] .
  \label{KW_action_chem}
\end{eqnarray}
From the viewpoint of the universality class, these two theories,
(\ref{naive_action}) and (\ref{KW_action_chem}), should belong to the
same class.

Here we remark the way of introducing chemical potential.
It was pointed out in~\cite{Hasenfratz:1983ba} that a naive form of the
chemical potential, $\mu\psi^\dag\psi=\mu\bar\psi\gamma_4\psi$, violates the Abelian gauge
invariance and requires a counter term to make thermodynamical
quantities finite.
On the other hand, in the KW fermion, the flavored chemical potential term is
introduced in this naive form.
It leads to necessity of tuning $\mu_3$ to deal with 
$\mathcal{O}(1/a)$ additive renormalization of chemical potential, 
as with the mass renormalization in the Wilson fermion.
This renormalization effect is relevant to the phase diagram in
the $(\mu_3$-$g^2)$ parameter plane, as discussed in~\cite{Misumi:2012uu}.

\section{Strong-coupling lattice QCD}

We study QCD phase diagram in the framework of the strong-coupling
lattice QCD with KW-type minimally doubled fermion.
We extend the strong-coupling analysis with this lattice
fermion~\cite{Kimura:2011ik} to the finite temperature and density
system~\cite{Nishida:2003uj,Fukushima:2003vi,Nishida:2003fb}.
The effective potential in terms of the meson
field is obtained by performing the 1-link integral in the strong coupling limit
$(g^2\to\infty)$, and then introducing auxiliary fields to eliminate the
4-point interactions.
In the case with KW fermion, we have to consider both of the scalar
$\sigma = \langle \bar\psi \psi \rangle$ and vector $\pi_4 = \langle
\bar \psi i \gamma_4 \psi \rangle$ condensates.
Identifying $\bar\psi\gamma_4=\psi^\dag$, the latter corresponds to the
imaginary density $i\langle\psi^\dag\psi\rangle$.
For the case with $\SU(N_c)$ gauge group and $d=D+1$ dimensions
in the finite temperature and density, we obtain the the following
effective potential,
\begin{equation}
 \mathcal{F}_{\rm eff}(\sigma,\pi_4;m,T,\mu,\mu_3,d_4)
  = \frac{N_c D}{4}
  \left(
   (1+r^2) \sigma^2 + (1-r^2) \pi_4^2
  \right)
  - N_c \log A - \frac{T}{4} \log
  \left(
   \sum_{n\in\Z} \det (Q_{n+i-j})_{1\le i,j\le N_c}
  \right).
  \label{eff_pot}
\end{equation}
In particular, the determinant part for $N_c=3$ is given by
\begin{eqnarray}
 &&
  \sum_{n \in \mathbb{Z}} \det\left(Q_{n+i-j}\right)_{1\le i,j\le N_c}
  \nonumber \\
   & = & 
   8 \left(
      1 + 12 \cosh^2 \frac{E}{T} + 8 \cosh^4 \frac{E}{T}
     \right)
   \left(
    15 - 60 \cosh^2 \frac{E}{T} + 160 \cosh^4 \frac{E}{T}
    -32 \cosh^6 \frac{E}{T} + 64 \cosh^8 \frac{E}{T}
   \right)
   \nonumber \\
 &&
  + 64 \cosh \frac{\mu_B}{T} \cosh \frac{E}{T}
  \left(
   -15 + 40 \cosh^2 \frac{E}{T} + 96 \cosh^4 \frac{E}{T}
   + 320 \cosh^8 \frac{E}{T}
  \right)
  \nonumber \\
 &&
  + 80 \cosh \frac{2\mu_B}{T}
  \left(
   1 + 6 \cosh^2 \frac{E}{T} + 24 \cosh^4 \frac{E}{T} 
   + 80 \cosh^6 \frac{E}{T}
  \right)
  \nonumber \\
 &&
  + 80 \cosh \frac{3\mu_B}{T}
  \cosh \frac{E}{T}
  \left(
   - 1 + \cosh^2 \frac{E}{T}
  \right)
  + 2 \cosh \frac{4\mu_B}{T},
\end{eqnarray}
with
\begin{equation}
 E = \mathrm{arcsinh} (B/A) , \quad
 A^2 = (1+d_4)^2 +
  \left(
   \mu_3 + D r - \frac{D}{2} (1-r^2) \pi_4
  \right)^2, \quad
 B = m + \frac{D}{2} (1+r^2) \sigma.
\end{equation}
Here the baryon chemical potential is defined as $\mu_B = 3 \mu$.
Remark that the next-leading order terms in $\mathcal{O}(1/\sqrt{D})$ are
omitted in the derivation.
See \cite{Misumi:2012ky} for the detailed calculation.

In the zero temperature case, we can
solve the equilibrium condition analytically.
For $D=3$ ($d=4$) with $m=0$ and $r=1$ the potential is given by
\begin{equation}
   \mathcal{F}_{\rm eff}(\sigma) = 
  \frac{9}{2}  \sigma^2 - \frac{3}{2} 
 \log \left( (1+d_4)^2 + (\mu_3+3)^2 \right)
 - \max
  \left\{
   3 \ \mathrm{arcsinh}
   \left(
    \frac{3 \sigma}{\sqrt{(1+d_4)^2+(\mu_3+3)^2}}
   \right),  \mu_B
  \right\} .
\end{equation}
In this case there are two local minima of the free energy as a function
of $\sigma$ at $\sigma=0$ and $\sigma=\sigma_0$.
%$\fen = -\mu_B - \frac{3}{2} \log \left( (1+d_4)^2 + (\mu_3+3)^2
%\right)$ at $\sigma = 0$ and $\fen = \frac{9}{2} \sigma^2 -
%\frac{3}{2} \log \left( (1+d_4)^2 + (\mu_3+3)^2 \right) - 3
%\mathrm{arcsinh} (\frac{3\sigma}{\sqrt{(1+d_4)^2+(\mu_3+3)^2}})$ at
%$\sigma = \sigma_0$.
This $\sigma_0$ can be determined by the gap equation, $\partial\fen / \partial
\sigma\Big|_{\sigma=\sigma_0}=0$,
%\begin{equation}
% \frac{\partial \fen}{\partial \sigma}\Bigg|_{\sigma=\sigma_0} = 0
%  \quad \longrightarrow \quad
% \sigma_0^2
%  \left[
%   1 + \frac{9\sigma_0^2}{(1+d_4)^2+(\mu_3+3)^2}
%  \right]
%  = \frac{1}{(1+d_4)^2+(\mu_3+3)^2} .
%\end{equation}
%Therefore we have
\begin{equation}
 \sigma_0^2 = \frac{(1+d_4)^2+(\mu_3+3)^2}{18}
  \left[
   \sqrt{1+\frac{36}{((1+d_4)^2+(\mu_3+3)^2)^2}} - 1
  \right] .
\end{equation}
Comparing these two local minima, we can show that the global minimum
changes from $\sigma = \sigma_0$ to $\sigma = 0$ at the critical
chemical potential as
\begin{equation}
 \mu_B^{\rm critical}(T=0) =    
  3 \, \mathrm{arcsinh}
   \left(
    \frac{3 \sigma_0}{\sqrt{(1+d_4)^2+(\mu_3+3)^2}}
   \right)
   - \frac{9}{2} \sigma_0^2 .
   \label{zero_temp_crit_chem_pot}
\end{equation}
This chiral phase transition is of 1st order because the order parameter
$\sigma$ changes discontinuously at this critical chemical potential.
We can also evaluate the baryon density $\rho_B = - \partial \fen /
\partial \mu_B$ at $T = 0$.
It turns out to be empty $\rho_B =0$ when $\mu_B < \mu_B^{\rm critical}$.
On the other hand, when $\mu_B > \mu_B^{\rm critical}$, it is saturated as
$\rho_B = 1$.

\begin{figure}[t]
 \begin{center}
  \includegraphics[width=16em]{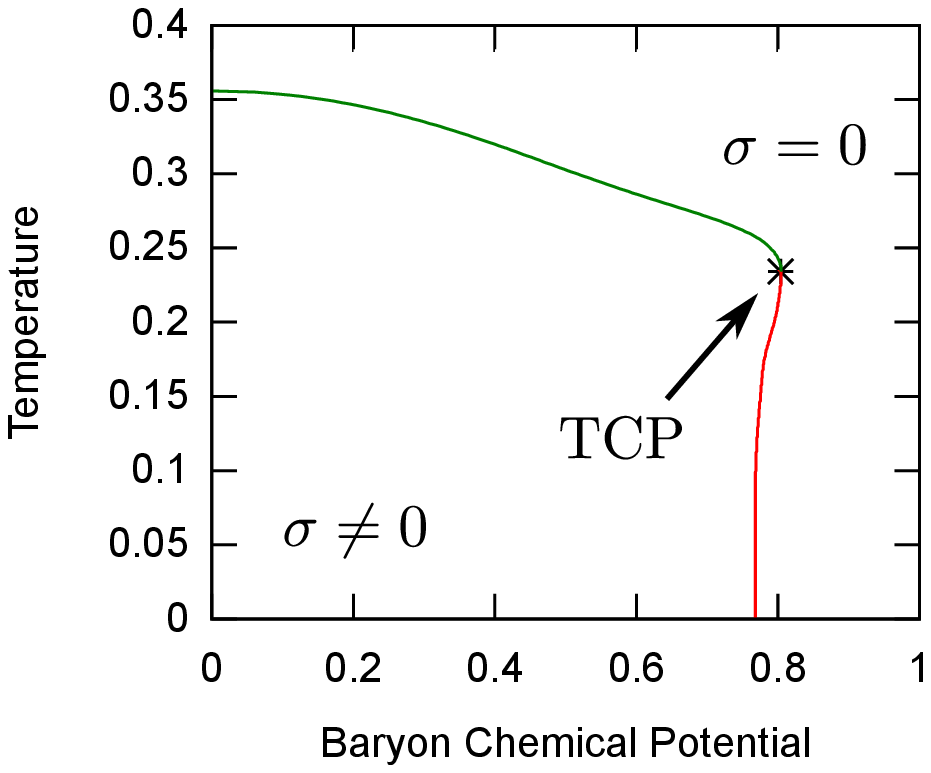} \qquad
  \includegraphics[width=19em]{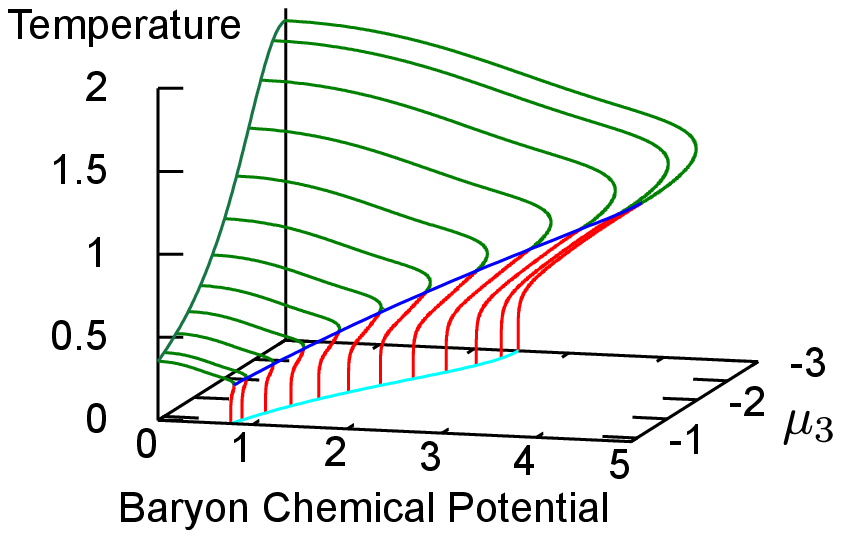}
 \end{center}
 \caption{(Left) Phase diagram for the chiral transition with $r=1$,
 $\mu_3=-0.9$ and $d_4=0$. Green and red lines show 2nd and 1st
 transition lines, respectively. The transition order is changed from
 2nd to 1st at the tricritical point $(\mu_B^{\rm tri},T^{\rm
 tri})=(0.804,0.234)$.
 (Right) Three-dimensional chiral phase diagram for $T$, $\mu_B$ and
 $\mu_3$ for $m=0$ where $\mu_{3}$ runs within half of the physical
 range $-3<\mu_3<\sqrt{32/7}-3$. 
 Green, red and purple lines show 2nd, 1st order transitions and
 tricritical point, respectively.}
 \label{pb_r1}
\end{figure}

\begin{figure}[t]
 \begin{center}
   \includegraphics[width=18em]{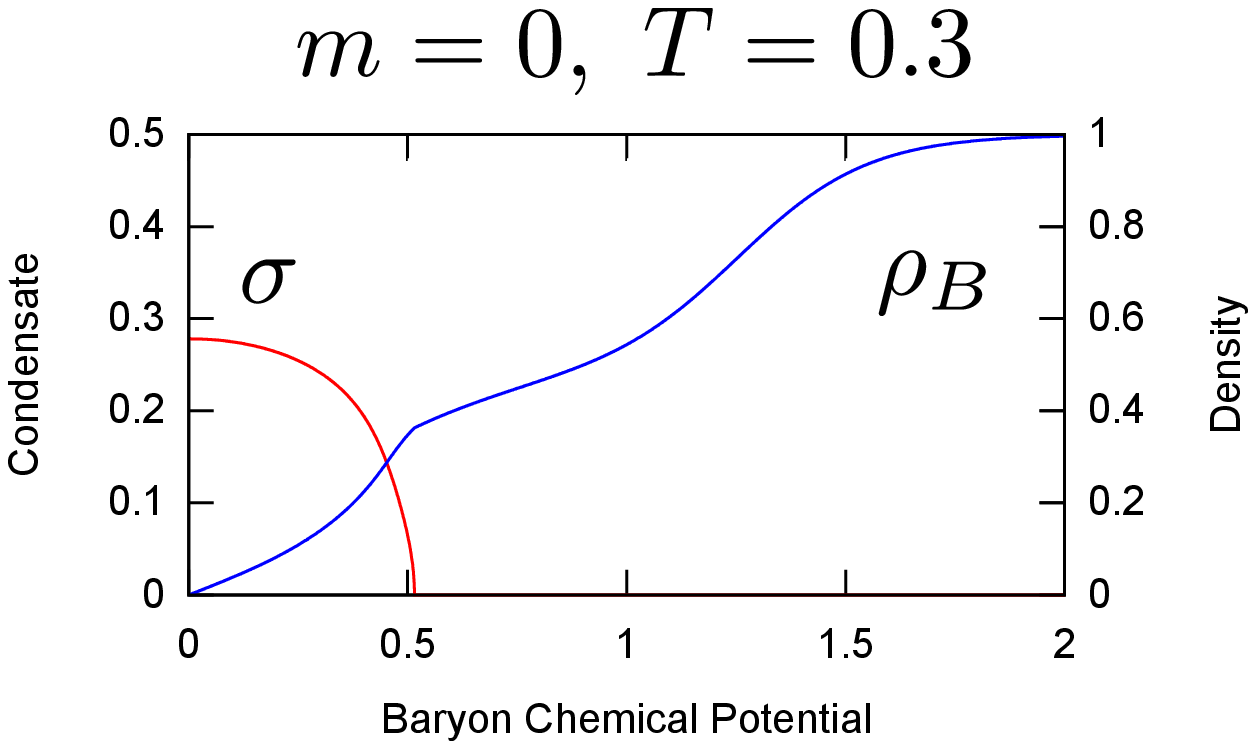} \quad
   \includegraphics[width=18em]{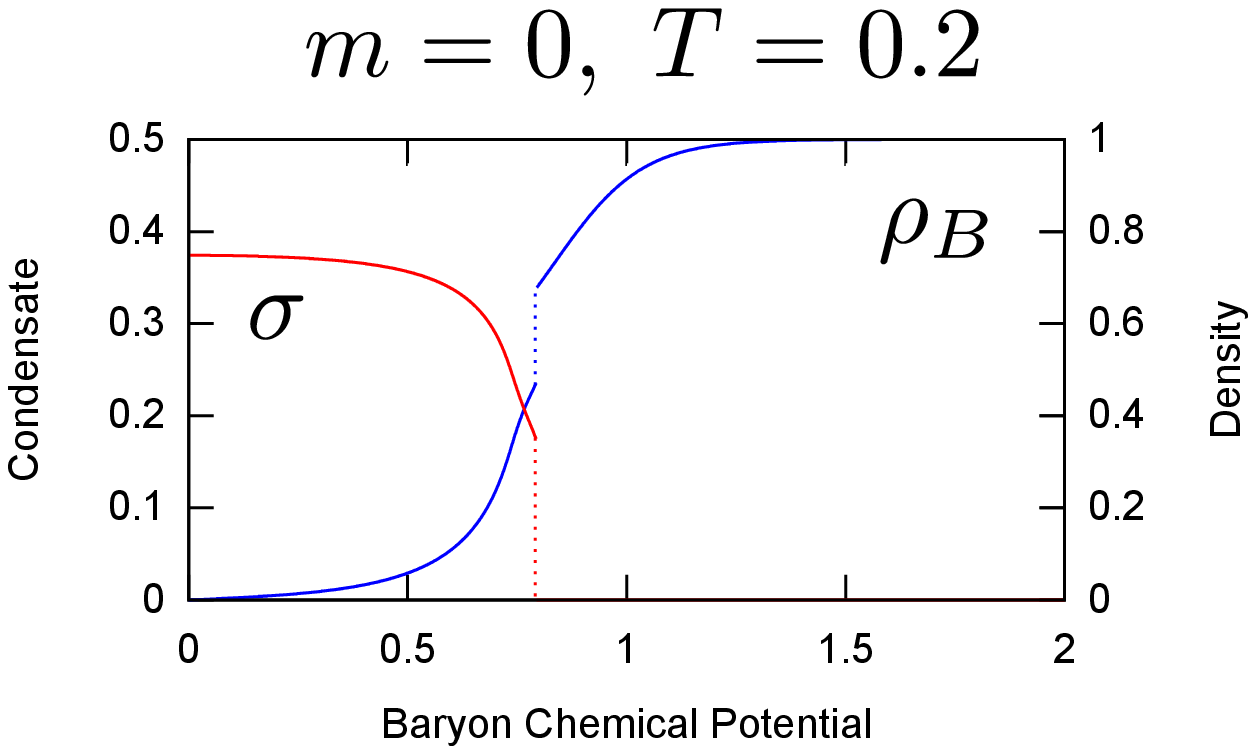} \\ \vspace{1.2em}
   \includegraphics[width=18em]{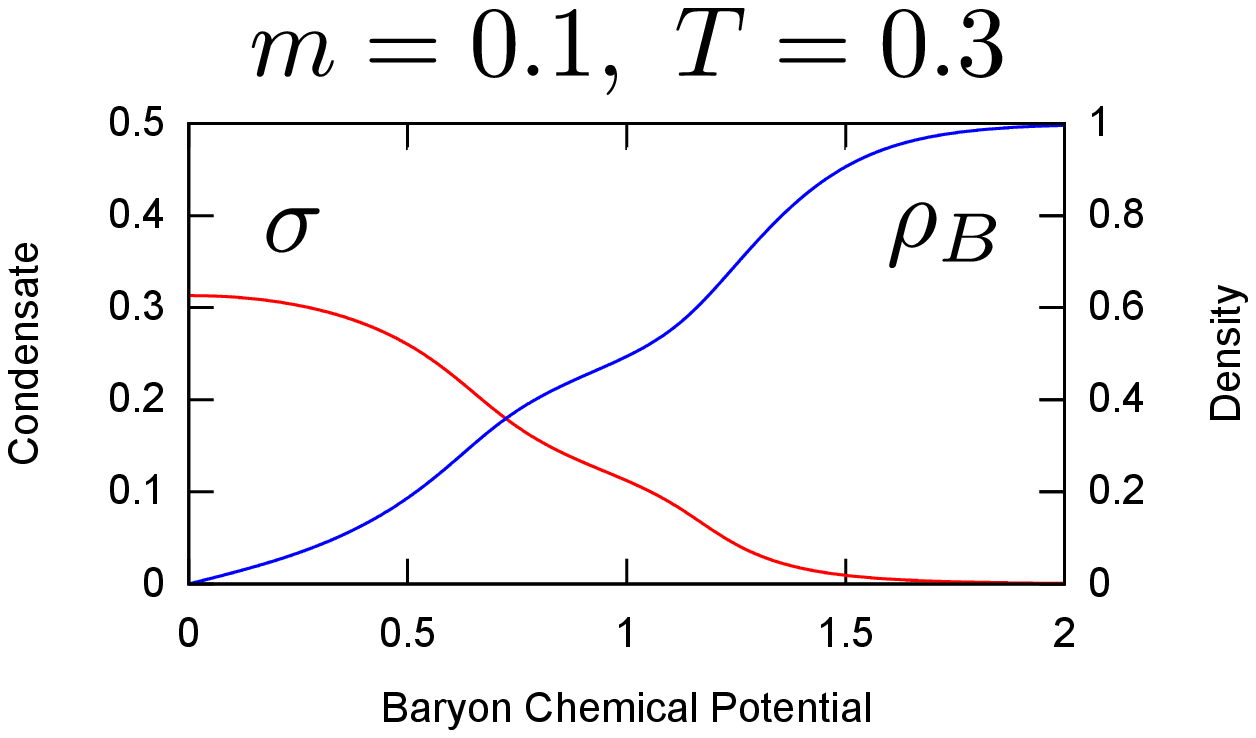} \quad
   \includegraphics[width=18em]{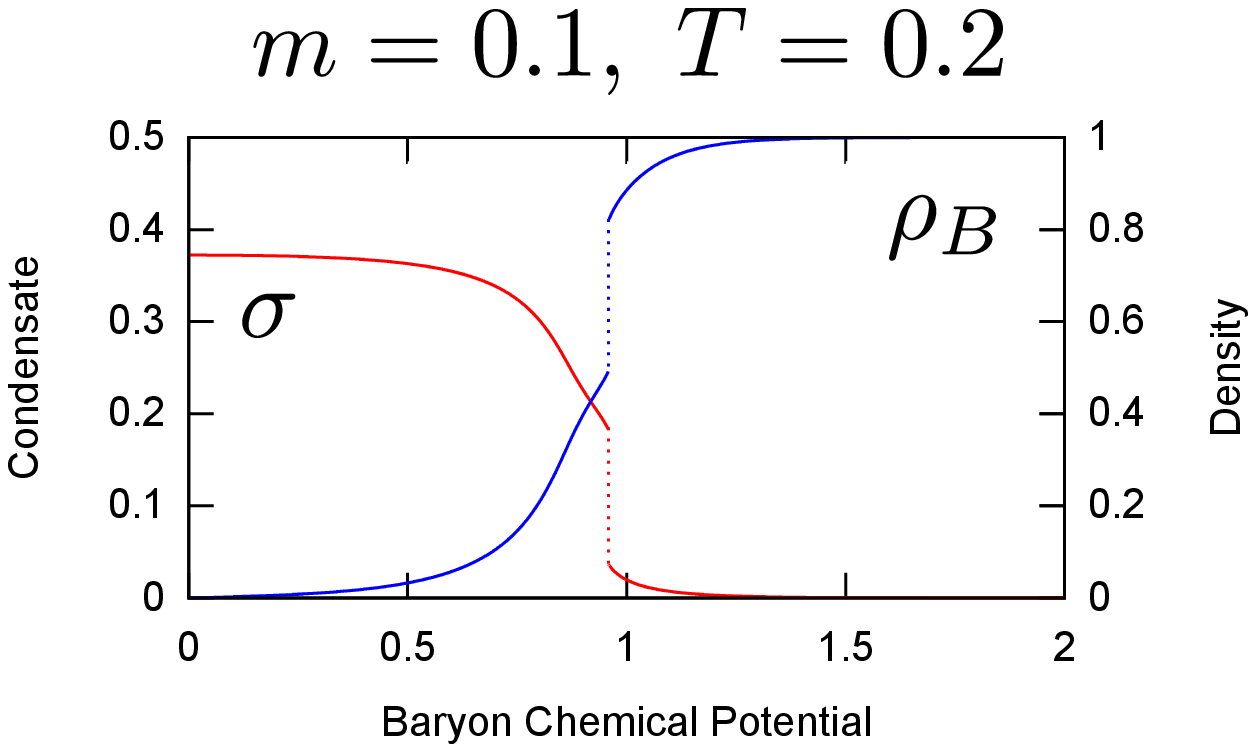}
  \end{center}
 \caption{Chiral condensate $\sigma$ and the baryon density $\rho_B$
 for (left) $T=0.3$ and (right) $T=0.2$ with $d_4=0$.
 Top and bottom panels show the massless $m=0$ and massive $m=0.1$ cases.
 There are 1st and 2nd phase transitions for $\sigma$.
 In the case of $m\not=0$, there appears the crossover behavior instead
 of the 2nd order transition.
 }
 \label{condensate_r1}
\end{figure}

We then discuss the phase diagram with respect to chiral symmetry.
We now concentrate on the case with $r=1$ for simplicity because the
effective potential (\ref{eff_pot}) is independent of $\pi_4$ in such a
case.
The 2nd order chiral phase boundary is given by the condition, such
that the coefficient of $\sigma^2$ in the effective potential
(\ref{eff_pot}) becomes zero.
When the order of the phase transition is changed from 2nd to 1st, the
coefficient of $\sigma^4$ as well as $\sigma^2$ should vanish.
The left panel of Fig.~\ref{pb_r1} shows the phase boundary of the
chiral transition with $r=1$, $\mu_{3}=-0.9$ and $m=0$ for $d_4=0$.
The counter term parameter is taken from the physical region
$-\sqrt{32/7}<\mu_3+3<\sqrt{32/7}$~\cite{Misumi:2012uu}.
The order of the phase transition is changed from 2nd to 1st at the
tricritical point $(\mu_B^{\rm tri},T^{\rm tri})=(0.804,0.234)$.
We also depict $\sigma$ condensate and the baryon density 
$\rho_B=-\partial \mathcal{F}_{\rm eff}/\partial \mu_{B}$
as functions of $\mu_{B}$ with several fixed $T$ in Fig.~\ref{condensate_r1}.
We find that there are 1st ($T<T^\mathrm{tri}$)
and 2nd ($T>T^\mathrm{tri}$) order phase transitions for $\sigma$, 
followed by the phase transition of the density $\rho_B$. 
For $m\not=0$, we can show that the crossover transition instead appears 
with the 2nd order critical point.

These results are qualitatively consistent with those with 
strong-coupling lattice QCD with staggered fermions,
while there are some quantitative differences.
For example, the KW phase diagram is suppressed in $T$ direction 
compared to that in staggered.
We here compare the ratio of the transition baryon chemical potential at $T=0$
to the critical temperature at $\mu_B=0$, $R^{0}=\mu_c(T=0)/T_c(\mu_B=0)$.
In staggered fermion, this ratio is
$R^0_\mathrm{st} \simeq 3 \times 0.56 / (5/3) \sim 1$~\cite{Fukushima:2003vi,Nishida:2003fb},
while $R^0_\mathrm{KW} \simeq 0.767 / 0.356 \sim 2.2$.
In the real world, this ratio is larger,
$R^0 \gtrsim M_N / 170~\mathrm{MeV} \sim 5.5$.
When the finite coupling and Polyakov loop effects are taken into account
for staggered fermion,
$T_c(\mu_B=0)$ decreases, $\mu_c(T=0)$ stays almost constant,
then $R^0$ value
increases~\cite{Miura:2009nu%,Miura:2008gd,Nakano:2009bf
}.
Larger $R^0$ with KW fermion in the strong coupling limit
may suggest smaller finite coupling corrections in the phase boundary.
Another interesting point is the location of the tricritical point.
In KW fermion, the ratio is $R^\mathrm{tri}_\mathrm{KW}=0.804/0.234 \simeq
3.4$,
while $R^\mathrm{tri}_\mathrm{st}=1.73/0.866 \simeq 2.0$
for unrooted staggered fermion~\cite{Fukushima:2003vi,Nishida:2003fb}.
It would be too brave to discuss this value,
but $R^\mathrm{tri}_\mathrm{KW}$ is consistent
with the recent Monte-Carlo
simulations (see references in \cite{Ohnishi:2011aa}),
%\cite{Fodor:2004nz,Ejiri:2003dc,Li:2011ee,deForcrand:2008vr,Ohnishi:2011aa}
which implies that the critical point does not exist
in the low baryon chemical potential region, $\mu_B/T \lesssim 3$.
These observations reveal
usefulness of KW fermion for research on QCD phase diagram. 

Apart from the phase transitions, 
the $\mu_B$ dependence of $\sigma$ and $\rho_B$ seems to
have some characteristics in Fig.~\ref{condensate_r1}.
At $T=0.3 > T^\mathrm{tri}$ with m=0, $\sigma$
and $\rho_B$ undergoes the 2nd-order phase transition 
at $\mu_B\simeq 0.5$,
and at a larger $\mu_B$ ($\mu_B\simeq 1.15$), 
increasing rate of $\rho_B$ as a function of $\mu_B$ becomes higher again.
At lower temperature, $T=0.2 < T^\mathrm{tri}$,
partial restoration of the chiral symmetry is seen
before the first order phase transition.
Since we have not taken care of the diquark condensate,
these continuous changes are not related to the color superconductor.
Other types of matter, such as quarkyonic matter~\cite{McLerran:2007qj},
partial chiral restored matter~\cite{Miura:2009nu}, or nuclear matter, 
may be related to the above characteristics.

In this report we focus on the case with $r=1$ and $d_4=0$ for
simplicity.
It is shown in~\cite{Misumi:2012ky} that effects of these parameters are
just quantitative.

\section{Summary}\label{sec:summary}

We have proposed a new framework for investigating
the two-flavor finite-($T$,$\mu$) QCD phase diagram. 
We have shown that the discrete symmetries of KW fermion 
strongly suggest its applicability to the in-medium lattice QCD. 
To support our idea, we study the strong-coupling lattice QCD
in the medium and derive the phase diagram of chiral symmetry 
for finite temperature and chemical potential.
We have obtained the phase diagram with 1st, 2nd-order and 
crossover critical lines, which is qualitatively in agreement to 
results from the model study.

\providecommand{\href}[2]{#2}\begingroup\raggedright\endgroup

\end{document}